\documentclass[sigconf]{aamas}

\usepackage{balance}
\usepackage{booktabs}
\usepackage{graphicx}
\usepackage{multirow}
\usepackage{algorithm}
\usepackage{algpseudocodex}
\usepackage{amsmath}
\usepackage{todonotes}

\makeatletter
\gdef\@copyrightpermission{
  \begin{minipage}{0.2\columnwidth}
   \href{https://creativecommons.org/licenses/by/4.0/}{\includegraphics[width=0.90\textwidth]{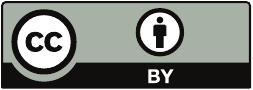}}
  \end{minipage}\hfill
  \begin{minipage}{0.8\columnwidth}
   \href{https://creativecommons.org/licenses/by/4.0/}{This work is licensed under a Creative Commons Attribution International 4.0 License.}
  \end{minipage}
  \vspace{5pt}
}
\makeatother

\setcopyright{ifaamas}
\acmConference[AAMAS 2027]{Proc.\@ of the 26th International Conference
on Autonomous Agents and Multiagent Systems (AAMAS 2027)}{3--7 May 2027}{Hanoi, Vietnam}{M.~Baldoni, F.~Fang, W.~Yeoh, N.~Yorke-Smith (eds.)}
\copyrightyear{2027}
\acmYear{2027}
\acmDOI{}
\acmISBN{}

\acmSubmissionID{2415}
\submissionType{Research Paper Track}

\title[Opponent Modeling under Partial Observability]{When Does Opponent Modeling Pay Off? Spatial Inference Stabilises Reinforcement Learning under Partial Observability}

\author{Marcel Petran}
\affiliation{\institution{Faculty of Mathematics and Physics, Charles University}\city{Prague}\country{Czech Republic}}
\author{Stefan Edelkamp}
\affiliation{\institution{Faculty of Mathematics and Physics, Charles University}\city{Prague}\country{Czech Republic}}

\begin{abstract}
Deep reinforcement learning agents in partially observable competitive
games must act on beliefs about an unseen opponent's intent. We study a
spatial opponent model: a transformer trained with hindsight-relabelled
opponent goals that outputs a subgoal heatmap from the opponent's observed
trajectory, fused early into a dueling double DQN. In symmetric
constant-sum \emph{foraging} games with partial observability, the module stabilises training.
Plain TD(0) DQN suffers degenerate seeds in three of four one-step
configurations, whereas
the opponent-model agent never collapses and, wherever training succeeds,
recovers the majority of the symmetric game value. Transfers zero-shot
to a \emph{foraging} opponent unseen by the learner, with a positive goal advantage over
the baseline, and loses this edge exactly where an information argument
predicts: against a
full-information tracking adversary whose intent is the learner itself,
and on resource-dense maps where both agents saturate. A blindfold
ablation, zeroing the heatmap at deployment, decomposes the advantage into
two mechanisms: training-time stabilisation, which dominates where the credit assignment is hard and survives blinding on the dense map, and
deployment-time inference, whose reliance shrinks as fog thins
on the sparse map. We conclude that
opponent-goal inference is a stability-and-transfer mechanism whose value
is governed by measurable boundary conditions: visibility (fog thickness), opponent
intent, and resource density.
\end{abstract}


\newcommand{\om}{\textsc{om}}
\newcommand{\classic}{\textsc{classic}}

\begin{document}

\maketitle

\section{Introduction}

\begin{figure}[t]
    \centering
    \includegraphics[width=0.48\linewidth]{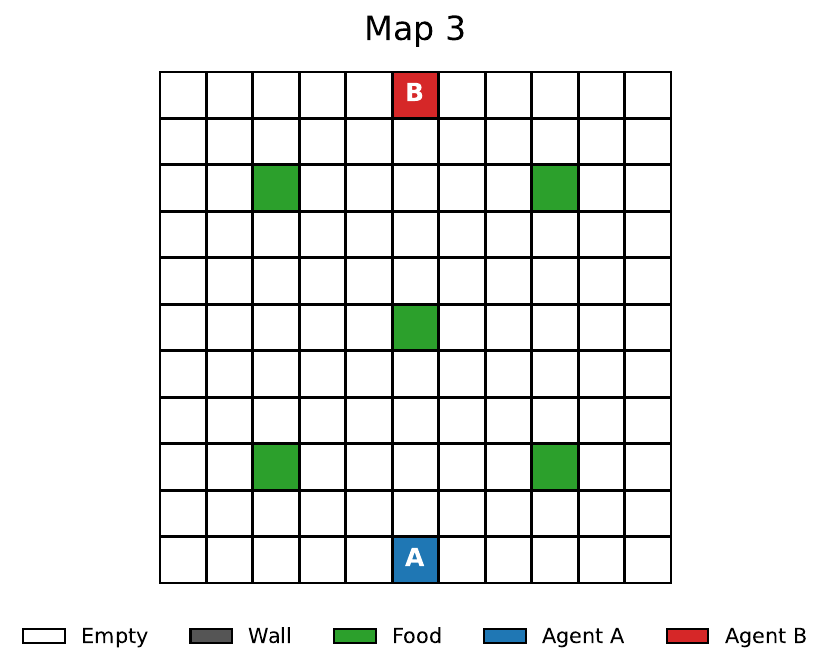}\hfill
    \includegraphics[width=0.48\linewidth]{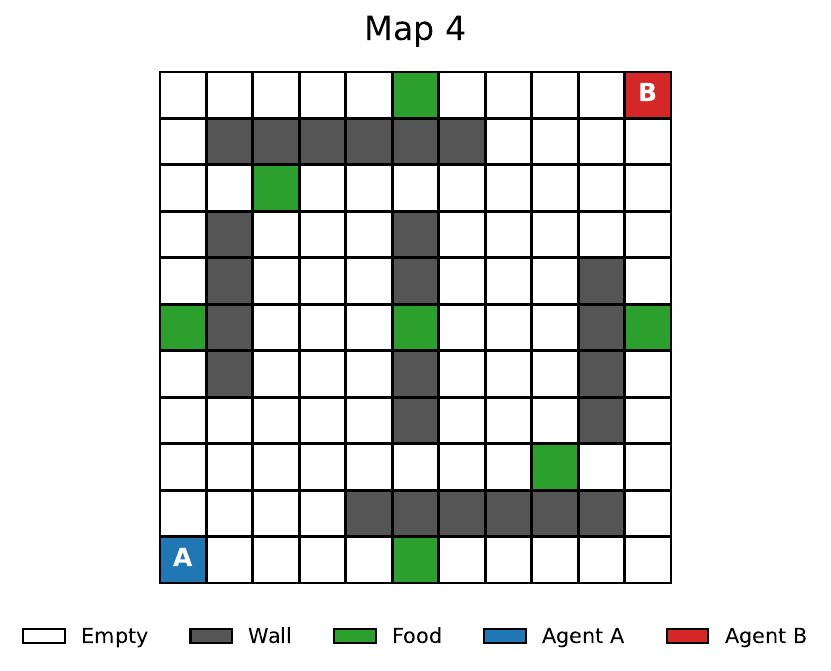}
    \caption{The two foraging maps ($11{\times}11$). \textsc{map3} (left):
    five densely placed central goals; \textsc{map4} (right): seven goals
    dispersed behind sectioned walls. Agents and foods spawn at fixed
    positions and compete for the resources.}
    \label{fig:maps}
\end{figure}

Competitive \emph{foraging} \cite{DBLP:conf/nips/ChristianosSA20} under fog (Figure \ref{fig:maps}) is a Partially Observable Stochastic Game (POSG) in which the pivotal latent is not where the opponent is, but where it is going. Classic DQN-style learners are hard to train in this regime: delayed sparse rewards combined with one-step temporal-difference learning (TD(0)) destabilise training. The ability to reason about others' behaviour can often be a defining factor between success and failure \cite{DBLP:conf/icml/RabinowitzPSZEB18, DBLP:journals/jair/NashedZ22, DBLP:conf/iclr/Al-ShedivatBBSM18, DBLP:conf/icml/HeB16, DBLP:conf/nips/PapoudakisCA21, DBLP:conf/nips/YuJL24, DBLP:journals/corr/abs-1802-09640}.
Relying solely on reactive policies is often insufficient. Instead, a prominent approach is to model the adversary directly, actively inferring their underlying strategies, intentions, or goals to anticipate future states. This approach is generally known as Opponent Modeling (OM). OM is usually evaluated only as ``beats the baseline,'' so we ask when it helps.
The spatial opponent model and its ESS formalism were introduced in prior
work~\cite{petran2026thesis} for Fully Observable Stochastic Games
(FOSG)~\cite{DBLP:conf/icml/Littman94}. Here we
adapt it to partial observability, replacing privileged inputs with the
fogged observation history and its first-order motion encodings, and fusing
the inferred heatmap alongside learned belief channels. Using this adapted
module, we contribute a boundary-condition analysis of when and why
opponent-goal inference pays off.

Our contributions are:
\textit{POSG instantiation of the spatial OM:} The opponent model
  of prior work~\cite{petran2026thesis} re-instantiated for fogged
  observation histories, with motion-difference channels, hindsight
  supervision, and early spatial fusion in physical coordinates.

\textit{Stability:} Across all training runs on two maps, the
  opponent-model agent never collapses, whereas plain DQN has degenerate
  seeds in three of four one-step configurations (every one-step
  configuration of the sparse map, and thick fog on the dense map).

\textit{In-distribution advantage: } Where training succeeds, the
  opponent model recovers 95--99.9\% of the symmetric game value against
  the training opponent (occasionally exceeding it by exploiting the
  suboptimal heuristic), versus 77--88\% for the baseline on \textsc{map4}
  and 52--107\% per seed on \textsc{map3}, with degenerate TD(0) training
  runs finishing as low as 47\%.

\textit{Generalisation across forager types:} The paired advantage
  transfers to a \emph{foraging} opponent unseen by the learner ($+0.31$--$0.52$) and,
  in well-trained scenarios, vanishes against a tracking adversary.

\textit{Boundary conditions:} The advantage disappears against
  opponents whose intent is the learner itself and saturates when resources
  are dense.

\textit{Mechanism:} A blindfold ablation decomposes the advantage
  into training-time stabilisation and deployment-time inference, and
  shows inference reliance scales with fog.
\section{Related Work}
Opponent modeling has a long history, from theory-of-mind-style
agent modeling~\cite{DBLP:conf/icml/RabinowitzPSZEB18, Premack_Woodruff_1978,DBLP:journals/ai/AlbrechtS18}
to game-specific predictors~\cite{DBLP:journals/entcom/BakkesSH09,DBLP:conf/icml/HeB16}.
Nashed and Zilberstein survey opponent-modeling methods and their
design axes~\cite{DBLP:journals/jair/NashedZ22}.
Closest to us, Yu et al.\ infer opponent \emph{subgoals} and integrate
them into policy learning~\cite{DBLP:conf/nips/YuJL24}; their subgoal
inference conditions on the full state, whereas we infer a spatial intent
distribution from partial observations, and ask when the
inference channel pays off rather than whether it improves a benchmark.
Goal and plan recognition share our prediction
target~\cite{DBLP:journals/ai/CharniakG93}; our hindsight-relabelled
transformer (Section~\ref{sec:her_generation}) is a lightweight online
variant of such methods, using realised opponent outcomes as supervision
in the spirit of HER~\cite{DBLP:conf/nips/AndrychowiczCRS17}.
Under partial observability, agent modeling has been tackled with
recurrent opponent embeddings~\cite{DBLP:conf/nips/PapoudakisCA21},
Bayes-adaptive meta-learners~\cite{DBLP:conf/iclr/ZintgrafSISGHW20}, and
policy-inference networks~\cite{DBLP:journals/corr/abs-1712-07893}; these
predict opponent \emph{policies or types}, not their spatial intent.
Our games are partially observable stochastic
games~\cite{DBLP:conf/aaai/HansenBZ04},
and our contributions are orthogonal to MARL algorithmics: we modify the
observation pipeline of a standard dueling
DDQN~\cite{DBLP:journals/nature/MnihKSRVBGRFOPB15,DBLP:conf/aaai/HasseltGS16,DBLP:conf/icml/WangSHHLF16}
and measure the effect on
training stability and transfer.

\section{Problem Formulation}
\label{sec:problem}
\begin{definition}[Partially Observable Stochastic Game]
A Partially Observable Stochastic Game (POSG) is a tuple
\begin{equation*}
\Gamma = \langle \mathcal{N}, \mathcal{S}, \mathcal{A}, \mathcal{T}, R_i, \mathcal{O}, Z_i, \gamma \rangle
\end{equation*}
where:
\begin{itemize}
    \item $\mathcal{N}$ is the set of $N$ agents.
    \item $\mathcal{S}$ is the global state space.
    \item $\mathcal{A} = \mathcal{A}_1 \times \cdots \times \mathcal{A}_N$ is the joint action space.
    \item $\mathcal{T}: \mathcal{S} \times \mathcal{A} \to \Delta(\mathcal{S})$ is the state transition function conditioned on joint actions, where $\Delta(\mathcal{S})$ denotes the probability simplex over the state space.
    \item $R_i: \mathcal{S} \times \mathcal{A} \to \mathbb{R}$, $i \in \mathcal{N}$, are the per-agent reward functions.
    \item $\mathcal{O}$ is the set of observation spaces $\mathcal{O} = \mathcal{O}_1 \times \cdots \times \mathcal{O}_N$.
    \item $Z_i: \mathcal{S} \to \mathcal{O}_i$, $i \in \mathcal{N}$, are the per-agent observation functions mapping the global state to agent $i$'s observation space.
    \item $\gamma \in [0, 1)$ is the discount factor.
\end{itemize}
\end{definition}

\paragraph{Foraging as a POSG}
We model competitive foraging under partial observability as a two-player constant-sum POSG $\Gamma = \langle \mathcal{N}, \mathcal{S},$ $\mathcal{A}, \mathcal{T},$ $R_i, \mathcal{O}, Z_i, \gamma \rangle$ with agents $\mathcal{N} = \{1,2\}$, $\mathcal{S}$ that represents the global state configurations that describe the positions of all agents, walls, foods, and the food's state (present or eaten). The agents share the movement action set $\mathcal{A} = \{ \text{North, South, East, West} \}$. $\mathcal{T}$ is deterministic, meaning that the movement actions update the agent's coordinates to the adjacent cell, agents automatically collect items upon entering a cell containing food and if agent $i \in \mathcal{N}$ takes an action that would result in a collision with a wall or stepping outside the boundaries of the map, agent $i$ stays at its current position. Each episode places both agents and a finite set of food goals on a grid. Agent $i$ receives a reward $r^i = 1$ if and only if it moves onto a cell containing food while the opponent is not present on that exact cell. If both agents enter the same cell containing food simultaneously, the food is split, yielding a reward of $r^i = 0.5$ for each agent. In all other scenarios, including empty movements and wall collisions, the step reward is $r^i = 0$. Agent $i$ does not observe the true state $s_t \in \mathcal{S}$: it receives a local observation $o^{(i)}_t = Z_i(s_t)$ restricted to a vision radius $v$ around
its position, with cells outside the radius masked. Agents are also unable to see through walls. All agents additionally maintain \emph{belief channels} $b_t$ (estimated food and opponent positions, updated on sightings), initialised
from the environment's opening state. The game is a partially observable multi-agent variant of the
travelling-salesman problem, making the grids minimal instances of the problem class we ultimately target.


\paragraph{Explicit Sequential Subgoal Games}
\label{sec:ess}
The ESS formalism and the spatial opponent model below were introduced in
prior work~\cite{petran2026thesis}; we summarise them for completeness and
describe our partial-observability adaptation. We now characterise the class of games for which inferring a spatial subgoal is the right prediction target. Many tasks are hierarchical: before winning, an agent must complete a sequence of intermediate objectives, or \emph{subgoals}. Predicting the opponent's full future state is unnecessarily noisy: the state carries contextual information that constrains movement but does not define the objective. We therefore decompose the state space as a Cartesian product
    $\mathcal{S} = \mathcal{P} \times \mathcal{Q}$
\noindent where $\mathcal{P}$ is the \emph{subgoal feature space}, the state variables whose values complete the atomic objective (e.g.\ an agent's coordinates), and $\mathcal{Q}$ is the \emph{context feature space}, variables that constrain behaviour but do not define the objective (walls, free space). Let $\rho: \mathcal{S} \to \mathcal{P}$, $\rho(s) = p$ for $s = (p, q)$, be the projection onto $\mathcal{P}$.

Let $\mathcal{S}^{+} = \{ s \in \mathcal{S} \mid R(s) > 0 \}$ be the rewarding states. We define the set of \emph{explicit subgoals} $\mathcal{G} = \rho[\mathcal{S}^{+}] \subseteq \mathcal{P}$. In the foraging POSG, $\mathcal{P}$ is the pair of agent coordinates, $\mathcal{Q}$ collects walls and remaining food, and $\mathcal{G}$ is the finite set of food coordinates.

\begin{definition}[ESS game]
A POSG is classified as an ESS game if and only if there exists a non-trivial decomposition of the state space such that $\mathcal{S} = \mathcal{P} \times \mathcal{Q}$, and the resulting set of distinct subgoals $\mathcal{G}$ is finite.
\end{definition}

\section{The Spatial Opponent Model under Partial Observability}
The Spatial Opponent Model (Spatial OM) was introduced for fully observable
games in prior work~\cite{petran2026thesis}. Relative to that instantiation,
partial observability requires three changes: (i) inference over a sliding
window of fogged observations rather than the full state, (ii)
motion-difference channels as the primary sighting signal, and (iii)
offline pre-training; in preliminary runs without it, the fusion channel
is ignored by the Q-network and does not recover. The prediction target is unchanged: the
adversary's physical destination as a dense heatmap over the grid, with
the ESS formalism restricting predictions to meaningful destinations.

\begin{figure}[H]
    \centering
    \includegraphics[width=0.9\linewidth]{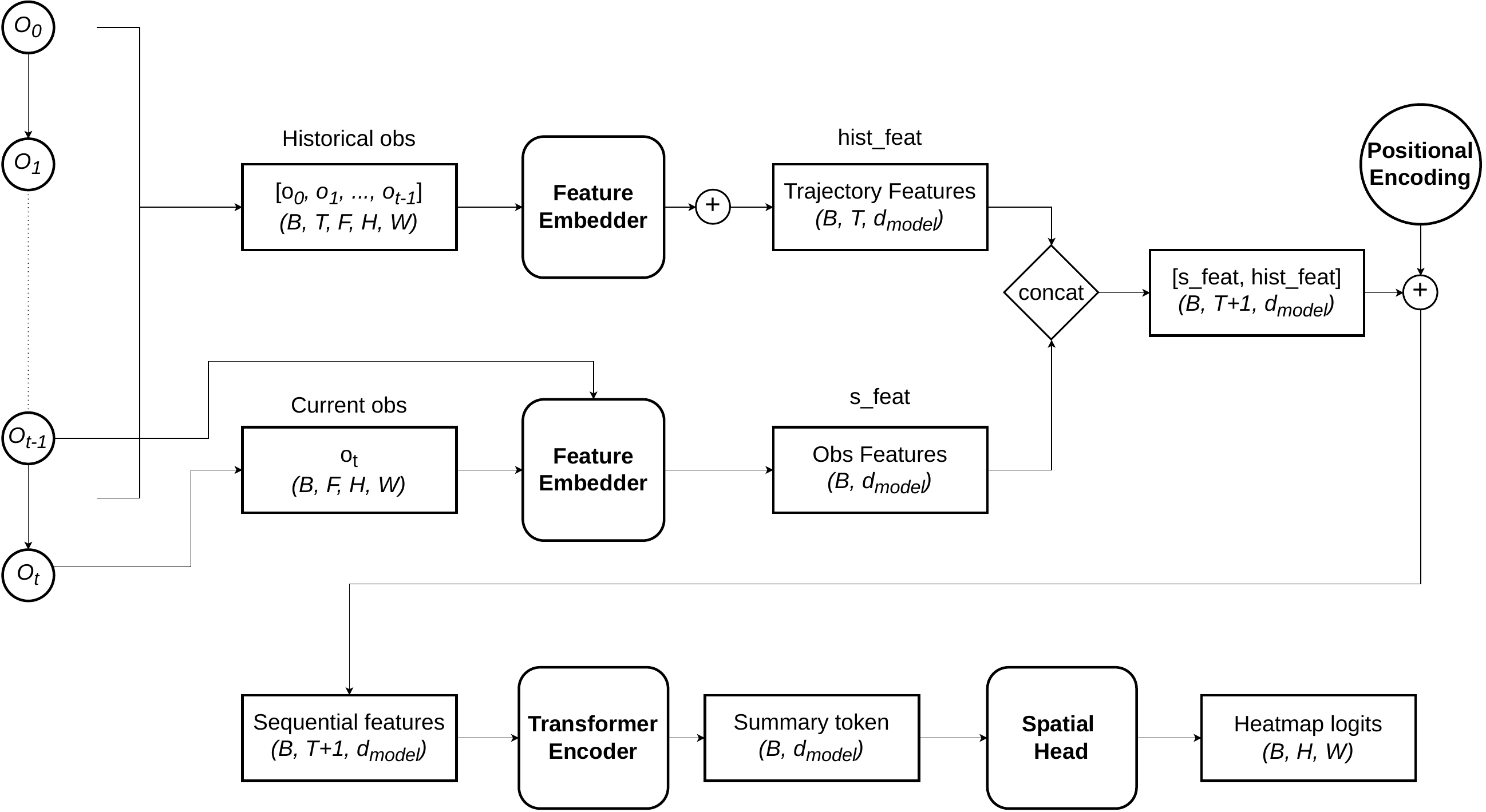}
    \caption[Transformer Architecture]{Overview of the Transformer-based Spatial Opponent Model. The Transformer Encoder processes the sequence of past observations, attending to the current observation to output a 2D heatmap representing the inferred opponent subgoals. The Feature Embedder is detailed in Figure \ref{fig:SOM_feature_embedder}. The diagram also shows expected tensor shapes where: $B$ is the batch dimension, $T$ is the trajectory dimension, $F$ is the environment Feature dimension, $H$ and $W$ are the environment Height and Width dimensions, and $d_{model}$ is the embedding output dimension.}
    \label{fig:SOM_architecture}
\end{figure}

\paragraph{Inference Model}
The core of our module is the inference model $p$, parameterised by $\phi$ (visualised in Figure \ref{fig:SOM_architecture}). It approximates the opponent's intended spatial destination at timestep $t$, given the learner's observation history and current observation, as a raw logit map $\hat{G}_{logits,t} \in \mathbb{R}^{H \times W}$ over the grid.

\begin{figure}[H]
    \centering
    \includegraphics[width=0.8\linewidth]{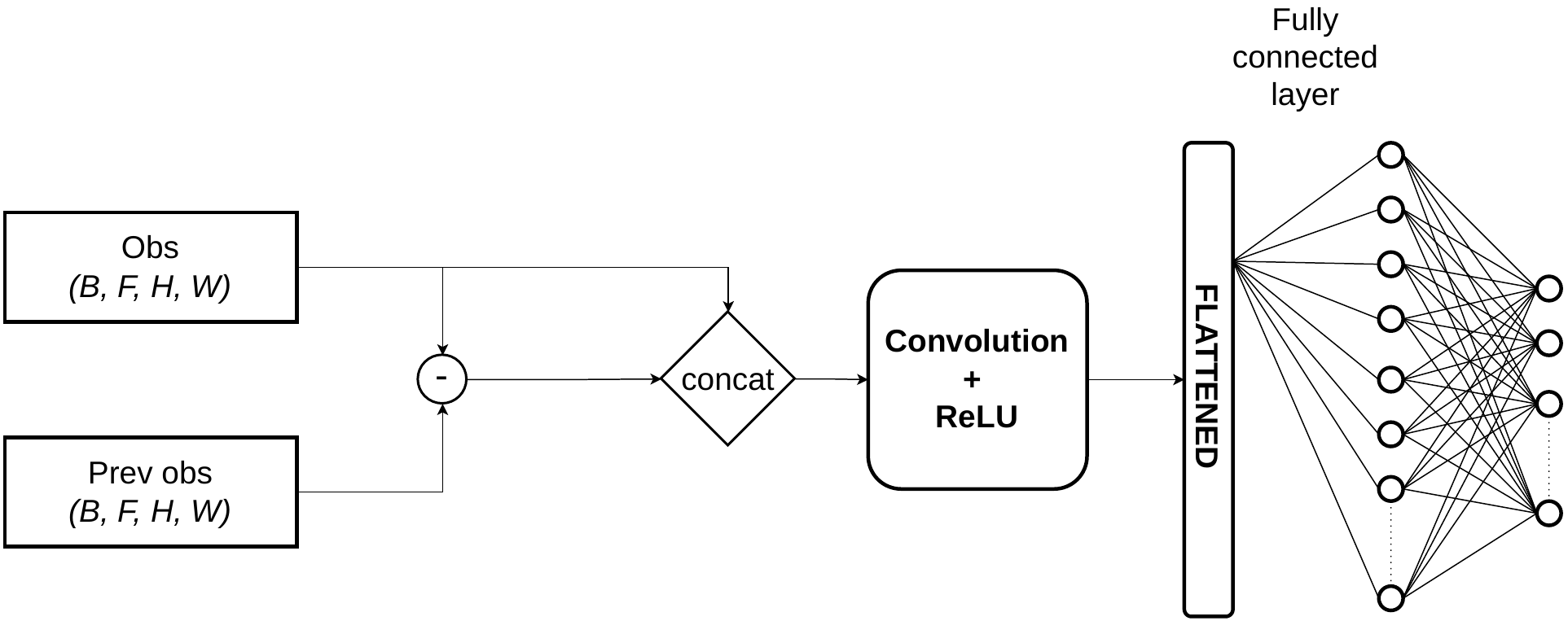}
    \caption[Feature Embedding]{Feature embedding process inside the Spatial Opponent Model.}
    \label{fig:SOM_feature_embedder}
\end{figure}

Let $\tau_t = (o_{t-T}, \dots, o_{t-1})$ be the learner's observation history over a fixed-length sliding window of maximum length $T$ (zero-padded for $t < T$). The architecture processes this sequence as follows:

\begin{enumerate}
    \item \textbf{Feature Embedding:} Each observation in $\tau_t$ and the current observation $o_t$ are processed through a shared feature embedding. As depicted in Figure \ref{fig:SOM_feature_embedder}, the spatial tensor $(2F \times H \times W)$, formed by concatenating the observation with its first-order motion encoding $o - o_{\mathrm{prev}}$, passes through a Convolutional layer with ReLU activation to extract local spatial features. The resulting feature map is downsampled, flattened and projected via a fully connected layer into a continuous vector of dimension $d_{model}$. The motion channels are the key partial-observability signal: a negative difference marks a cell that just \emph{disappeared under fog} or the opponent leaving the vision radius.
    \item \textbf{Transformer Encoder:} A multi-layer Transformer Encoder with standard positional encodings processes the sequence; a boolean key-padding mask prevents self-attention from attending to unobserved history.
    \item \textbf{Spatial Head:} To generate the prediction, the model extracts the ``summary token'' (the first output token, corresponding to the current observation $o_t$) and passes it through a shallow MLP that projects the $d_{model}$ vector back into the spatial dimensions $(H \times W)$, yielding the heatmap logits $\hat{G}_{logits, t} \in \mathbb{R}^{H \times W}$.
\end{enumerate}

We define the probability of the opponent targeting a specific coordinate $(i, j) \in H \times W$ as the output of the inference model $p_{\phi}$. The Spatial Head produces a raw logit map, $\hat{G}_{logits, t} \in \mathbb{R}^{H \times W}$. To ensure the final output represents a valid spatial probability distribution, we apply a Softmax activation function. This produces the predicted spatial intent heatmap $\hat{G}_{t} \in [0, 1]^{H \times W}$, where each element represents the probability of the opponent targeting a specific coordinate $(i, j)$ conditioned on the learner's current observation $o_t$ and observation history $\tau_t$:

\begin{equation}
\label{eq:heatmap}
    \hat{G}_{t}^{(i,j)} = p_{\phi}(g_t = (i,j) \mathrel{|} o_t, \tau_t) = \frac{\exp(\hat{G}_{logits, t}^{(i,j)})}{\sum_{x=1}^{H} \sum_{y=1}^{W} \exp(\hat{G}_{logits, t}^{(x,y)})}
\end{equation}

\noindent where the resulting heatmap $\hat{G}_t$ is constrained such that the total probability mass sums up to 1:
\begin{equation}
    \sum_{i=1}^{H} \sum_{j=1}^{W} \hat{G}_t^{(i,j)} = 1
\end{equation}

By treating the opponent's observable past as a sequence of tokens, the Transformer leverages self-attention to capture long-range temporal dependencies between historical movements and current intent. Training the model, however, requires ground-truth labels that are not directly observable in a decentralised multi-agent setting.

\paragraph{Hindsight Target Generation}
\label{sec:her_generation}
To optimise the Spatial OM, the network requires spatial ground-truth labels.
Because the opponent's internal decision-making is not always accessible to the
learner during an episode, we generate labels post-hoc with Hindsight Target
Generation (strongly related to Hindsight Experience Replay
\cite{DBLP:conf/nips/AndrychowiczCRS17}).

At the conclusion of an episode, the trajectory is analysed backwards to determine the opponent's achieved subgoals. If the opponent successfully reaches a food at timestep $t+k$, that coordinate $(x, y)$ becomes the ground-truth subgoal for the preceding sequence; if the opponent fails to reach a definitive objective before the episode ends, their final occupied position becomes the fallback target. To account for spatial uncertainty and avoid overly sparse gradients, we transform this single coordinate into a continuous target intent heatmap $G_{target, t} \in [0, 1]^{H \times W}$ by applying a 2D Gaussian smoothing kernel with variance $\sigma^2$ centred at the target coordinate:

\begin{equation}
    G_{target, t}^{(i, j)} = \exp\left( -\frac{(i - x)^2 + (j - y)^2}{2\sigma^2} \right)
\end{equation}

This smoothing provides a dense learning signal, penalising ``near-misses'' less severely than spatially distant errors. The systematic process for extracting these labels from raw trajectories is detailed in Algorithm \ref{alg:her_labeling}.

\begin{algorithm}[t]
\caption{Hindsight Subgoal Labelling with Gaussian Smoothing}
\label{alg:her_labeling}
\begin{algorithmic}[1]
\Require
\Statex Logged episode $E = [(o_t, r^{\text{opp}}_t, s^{g}_{t}, s^{g}_{t+1})]_{t=1}^{T_{ep}}$: the learner's observations $o_t$, the opponent's rewards $r^{\text{opp}}_t$, and the pre-step and post-step privileged global states $s^{g}_{t}$ and $s^{g}_{t+1}$ (used for labelling only)
\Statex Grid dimensions $H \times W$, Smoothing variance $\sigma^2$
\State $g_{true} \leftarrow \text{None}$ \Comment{Current true goal coordinate $(x,y)$}
\State $G_{next} \leftarrow \mathbf{0}_{H \times W}$ \Comment{Target map for the next state}

\Statex \Comment{Fallback goal for unsuccessful or truncated episodes}
\If{$E$ is not empty}
    \State $(o_{T_{ep}}, r^{\text{opp}}_{T_{ep}}, s^{g}_{T_{ep}}, s^{g}_{T_{ep}+1}) \leftarrow E[T_{ep}]$
    \If{$r^{\text{opp}}_T = 0$}
        \State $g_{true} \leftarrow \text{ExtractOpponentPosition}(s^{g}_{T})$
    \EndIf
\EndIf

\Statex \Comment{Backwards Relabelling Trajectory}
\For{$t = T_{ep}$ \textbf{downto} $1$}
    \State $(o_t, r^{\text{opp}}_t, s^{g}_{t}, s^{g}_{t+1}) \leftarrow E[t]$

    \If{$r^{\text{opp}}_t > 0$} \Comment{Opponent achieved an objective (e.g., reached food)}
        \State $g_{true} \leftarrow \text{ExtractOpponentPosition}(s^{g}_{t+1})$
    \EndIf

    \State $G_{target, t} \leftarrow \mathbf{0}_{H \times W}$
    \If{$g_{true} \neq \text{None}$}
        \State $G_{target, t}[g_{true}] \leftarrow 1.0$ \Comment{Assign one-hot target}
        \State $G_{target, t} \leftarrow \text{GaussianFilter}(G_{target, t}, \sigma^2)$ \Comment{Generate dense soft-target map}
    \EndIf

    \State Label $G_{target, t}$ to transition $t$ as \texttt{true\_goal\_map}
    \State Label $G_{next}$ to transition $t$ as \texttt{true\_goal\_map\_next}
    \State $G_{next} \leftarrow G_{target, t}$ \Comment{Carry target map backward}
\EndFor

\State \Return Labelled episode $E$
\end{algorithmic}
\end{algorithm}

\paragraph{Training Objective}
Because $G_{target, t}$ is a continuous Gaussian density, we normalise it to
a valid probability distribution (summing to $1$ across the grid) and
optimise the network against it with the KL Divergence objective:

\begin{equation}
\begin{split}
    \mathcal{L}_{\mathrm{OM}}(\phi)
    &= \mathcal{D}_{KL}(G_{target, t} \parallel \hat{G}_t) \\
    &= \sum_{i=1}^{H} \sum_{j=1}^{W} G_{target, t}^{(i,j)}
       \bigl( \log G_{target, t}^{(i,j)}
       - \log \hat{G}_t^{(i,j)} \bigr)
\end{split}
\end{equation}

This encourages the Transformer to capture the regional spatial intent rather than memorising discrete coordinate points.

\paragraph{Feature fusion}
The Q-function is conditioned on the inferred intent: $Q(o_t, a, \hat{G}_t)$.

\begin{figure}[H]
    \centering
    \includegraphics[width=0.9\linewidth]{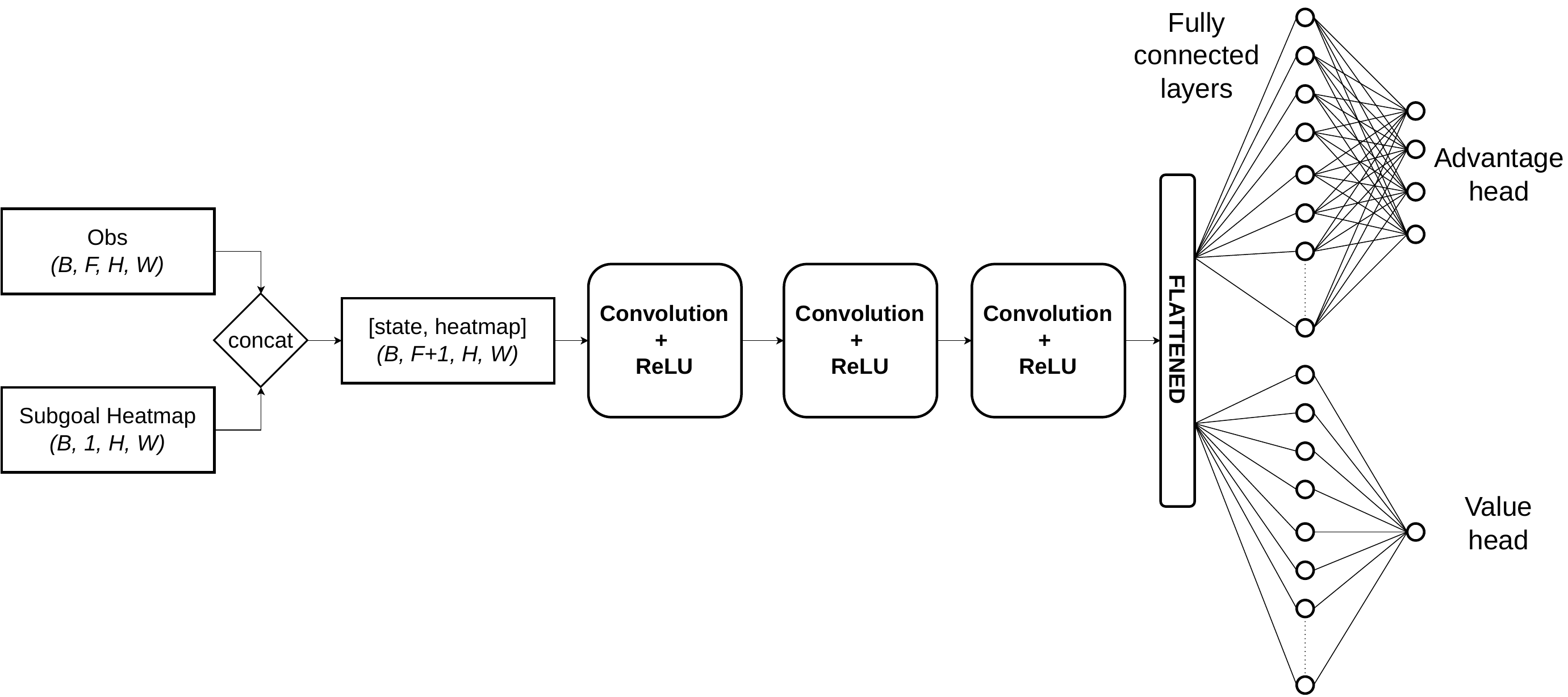}
    \caption[OM Dueling Q-Network]{OM Dueling Q-Network. The observation and belief channels are concatenated with the inferred heatmap and processed by the CNN; Advantage and Value heads compute the final Q-values by Equation \ref{eq:dueling_q}.}
    \label{fig:Q_net}
\end{figure}

A low-dimensional latent summary of the opponent's intent is easily drowned out by the far larger flattened spatial embedding of the Q-network's CNN, so the intent signal is kept in physical coordinates: in an ESS game the opponent's intent is tied to the grid's physical layout. The Spatial Head therefore projects intent back into a heatmap, and the resulting $\hat{G}_t$ (Eq.~\ref{eq:heatmap}) is concatenated as an additional channel of the network input (early spatial fusion) as depicted in Figure \ref{fig:Q_net}. The Q-network's CNN thus sees the opponent's likely destination as a heatmap overlaid on walls, food, and the agent's own position, and learns their geometric relationships directly.

Let $o_t \in \mathbb{R}^{F \times H \times W}$ be the learner's observation with $F$ feature channels, and $b_t \in \mathbb{R}^{M \times H \times W}$ with $M$ belief feature channels. This belief map is shared by all compared architectures. The inferred heatmap is concatenated with both along the feature dimension:

\begin{equation}
    x_{t} = [o_t; b_t; \hat{G}_t] \quad \in \mathbb{R}^{(F+M+1) \times H \times W}
\end{equation}

\noindent The embedding $\mathbf{h}_{cnn} = f_{\text{CNN}}(x_{t})$ is processed by the dueling heads \cite{DBLP:conf/icml/WangSHHLF16}, which separate state value ($V(\cdot)$) from per-action advantages ($A(\cdot, \cdot)$). The final Q-value is computed as:

\begin{equation}
\label{eq:dueling_q}
    Q(o_t, a, \hat{G}_t) = V(\mathbf{h}_{cnn}) + A(\mathbf{h}_{cnn}, a) - \frac{1}{|\mathcal{A}|} \sum_{a'} A(\mathbf{h}_{cnn}, a')
\end{equation}

\noindent The Q-network is optimised with the double-DQN
update~\cite{DBLP:conf/aaai/HasseltGS16}: the online network selects the
greedy next action and the Polyak-averaged target network $Q_{\theta^-}$
evaluates it, with target
\begin{equation}
\label{eq:ddqn_target}
    y_t = r_t + \gamma\, Q_{\theta^-}\bigl(o_{t+1}, \operatorname*{argmax}_{a'} Q_\theta(o_{t+1}, a', \hat{G}_{t+1}), \hat{G}_{t+1}\bigr)
\end{equation}
and loss $\text{Huber}(y_t - Q_\theta(o_t, a_t, \hat{G}_t))$ \cite{10.1214/aoms/1177703732}
(Algorithm~\ref{alg:om_ddqn}).

\paragraph{Pre-training the opponent model}
\label{sec:pretrain}
If the Spatial OM and the Q-network are trained concurrently from random initialisation, the early heatmaps are indistinguishable from uniform noise; the Q-network's CNN quickly learns to assign near-zero weights to the heatmap channel, and once ignored the channel rarely recovers. We therefore pre-train $p_\phi$ offline before joint RL training. A supervised dataset of trajectories is collected by simulating games between heuristic agents (various pairings of the \textsc{simple} and \textsc{greedy} profiles, Section \ref{sec:setup}) and labelled via hindsight labelling (Algorithm \ref{alg:her_labeling}). This grants the Spatial OM a foundational understanding of the grid's topology, agent movement, and the environment's rules, so that when joint OM-DDQN training begins (Algorithm \ref{alg:om_ddqn}) the Q-network receives structurally meaningful intent predictions from the first episode and integrates the heatmap channel during value estimation from the outset.

\begin{algorithm}[t]
\caption{DDQN with the Spatial OM (OM-DDQN) using HER}
\label{alg:om_ddqn}
\begin{algorithmic}[1]
\Statex \textbf{Initialise:}
    \State \quad Replay Buffer $\mathcal{D}_{RL}$
    \State \quad Agent Network $Q_\theta$, Target Network $Q_{\theta^-}$
    \State \quad Spatial Opponent Model $p_\phi$

\Statex \textbf{Pre-train:}
    \State \quad Train $p_\phi$ on offline heuristic dataset labelled via HER.
\Statex \textbf{Collect data:}
    \For{episode $= 1$ to $N_{\mathrm{ep}}$}
        \State Initialise environment, observe $o_0$, set history $\tau_0 = \emptyset$, and local episode buffer
        \For{step $t = 0$ to $T_{ep}$}
            \State \textbf{Inference:}
            \State \quad Predict subgoal heatmap: $\hat{G}_t = p_\phi(o_t, \tau_t)$
            \State \textbf{Action:}
            \State \quad Select $a_t$ using Boltzmann exploration on $Q_\theta(o_t, \cdot, \hat{G}_t)$
            \State \textbf{Step:}
            \State \quad Execute $a_t$, observe reward $r_t$ and next observation $o_{t+1}$
            \State \quad Update history: $\tau_{t+1} \leftarrow \text{append}(\tau_t, o_t)$ \Comment{slide window of length $T$}
            \State \quad Store step data $(o_t, a_t, r_t, o_{t+1}, \tau_t, \hat{G}_t)$ in local episode buffer
        \EndFor
        \State \textbf{Hindsight Labelling:}
            \State Label the logged episode as described in Algorithm \ref{alg:her_labeling}
            \State Transfer local buffer to global Replay Buffer $\mathcal{D}_{RL}$
        \Statex
        \State \textbf{Training Phase:}
            \State Sample batch of transitions and sequences from $\mathcal{D}_{RL}$
            \Statex \Comment{Train OM Dueling Q-Network}
            \State Generate next step heatmaps $\hat{G}_{t+1}$ using the Spatial OM inference $p_\phi(o_{t+1}, \tau_{t+1})$
            \State Select next actions $a'_{max} = \operatorname*{argmax}_{a'} Q_\theta(o_{t+1}, a', \hat{G}_{t+1})$
            \State Calculate target $y = r_t + \gamma Q_{\theta^-}(o_{t+1}, a'_{max}, \hat{G}_{t+1})$
            \State Compute Loss $\mathcal{L}_{DDQN} = \text{Huber}(y - Q_\theta(o_t, a_t, \hat{G}_t))$
            \State Update $\theta$ via gradient descent
            \State Update target network (Polyak): $\theta^- \leftarrow \beta\, \theta + (1-\beta)\, \theta^-$
            \Statex \Comment{Train the Spatial OM}
            \State Compute $\mathcal{L}_{OM}(\phi)$ using KL Divergence against normalised soft targets $G_{target, t}$
            \State Update $\phi$ via gradient descent
    \EndFor
\end{algorithmic}
\end{algorithm}

\begin{figure}[t]
    \centering
    \includegraphics[width=0.9\linewidth]{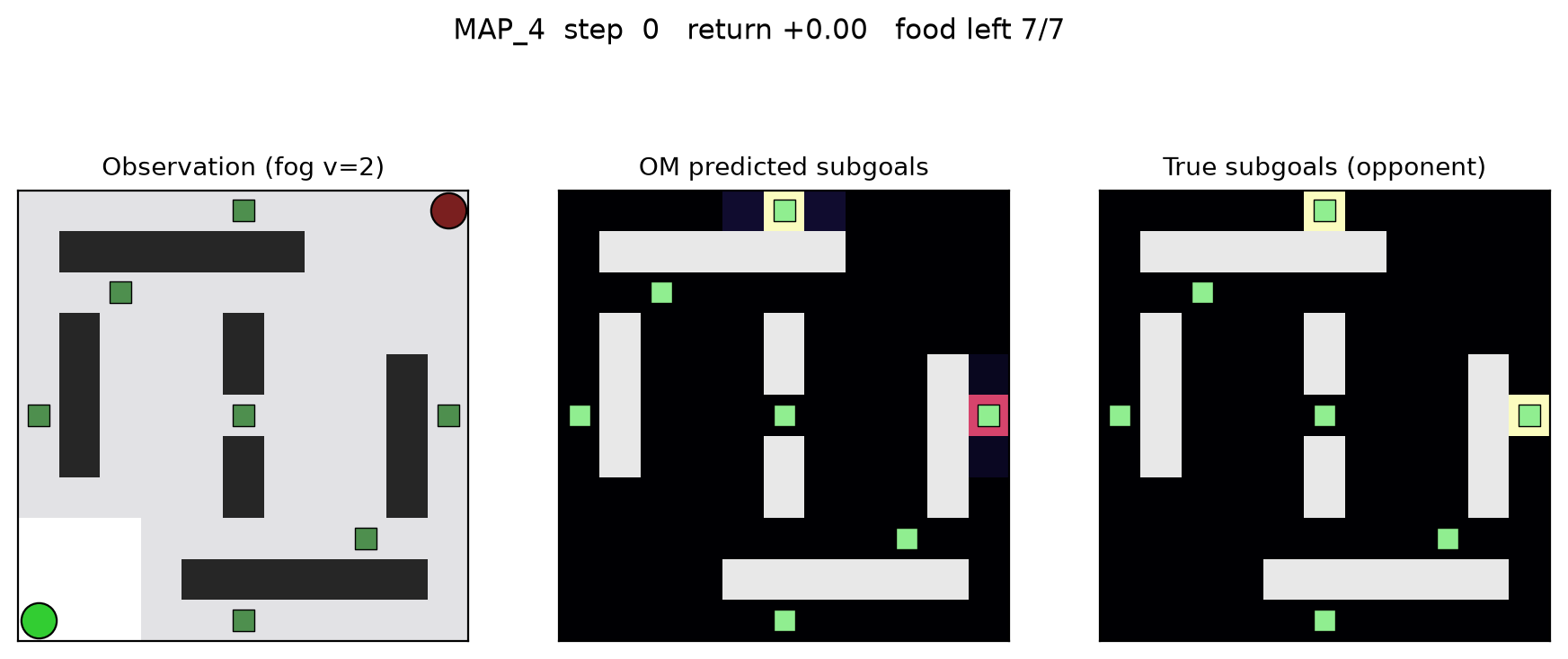}
    \caption[OM predictions]{Example of Spatial OM predictions on \textsc{map4} (vision radius 2). Left: the learner's observation under fog; the learner (green dot) sees only its surroundings, while the opponent's true position (dark red when obscured by fog) and the food (green squares) are shown for reference. Middle: the OM's predicted subgoal distribution fed to the Q-network. Right: the opponent's true subgoal distribution, that is not accessible during match. The OM concentrates mass on the opponent's actual target even when the opponent is outside the learner's vision radius.}
\end{figure}

\section{Experimental Setup}
\label{sec:setup}

\paragraph{Environment.} Foraging under partial observability (Section \ref{sec:problem}) played on two symmetric
$11{\times}11$ grids (Figure~\ref{fig:maps}): \textsc{map3} (dense central
food, 5 goals) and \textsc{map4} (sectioned walls, dispersed food, 7 goals),
with fixed spawn and goal positions and vision radius $v \in \{2,5\}$ (fog
thickness). The symmetry and game rules ensure that each agent's game value is half of the total food. Greedy vs greedy play shows
this empirically (\textsc{map3}: 2.50/2.50; \textsc{map4}: 3.42/3.42).

\paragraph{Opponents.} All opponents share the learner's action space and
movement rules and act on their own partial observations (same vision
radius). \textsc{simple} commits to a random believed food and replans
when it is consumed or seen as collected. \textsc{greedy} targets its nearest believed food
and switches away whenever it believes the learner is closer. \textsc{oracle-stalker} (also denoted as ``tracker'') sees the true state regardless of fog
and races to the food nearest the learner that it can still reach first, loitering there to deny collection.

\paragraph{Training.} Dueling DQN, $n$-step returns $n \in \{1,3\}$~\cite{DBLP:books/lib/SuttonB2018},
belief-map prior $p \in \{0,1\}$ (belief map off/on), 60 epochs, 3 seeds per configuration,
against the \textsc{greedy} opponent. Training used a single Tesla V100-SXM2-32GB GPU per run, taking about 7 hours plus per-epoch evaluation (\classic{} and \om{} combined). Zero-shot evaluations ran on an Apple M3 Pro (12-core CPU,
18\,GB RAM). Table~\ref{tab:hyper} lists the shared
hyperparameters.

\begin{table}[t]
\centering
\caption{Shared training hyperparameters and network sizes
(both learning agents).}
\label{tab:hyper}
\footnotesize
\begin{tabular}{@{}lp{0.52\columnwidth}@{}}
\toprule
Learning rate & Adam; $3{\cdot}10^{-4}$ (Q), $1{\cdot}10^{-4}$ (OM) \\
Discount $\gamma$ / soft update $\tau_{soft}$ & 0.985 / 0.001 \\
Batch size / train frequency & 128 / every 4 steps \\
Replay capacity & 450{,}000 \\
Episodes & 30{,}000 (60 epochs$\times$500 episodes), max 50 steps \\
Boltzmann $\tau$ & $1.5 \to 0.05$ over 650k steps \\
Label smoothing $\sigma^2$ & $2.0 \to 0.5$ over 150k steps \\
OM pre-training & 20 epochs heuristic self-play \\
Q-net & $3{\times}$Conv$3{\times}3$ ($C_{in}{\to}32{\to}64{\to}64$),
  dueling heads 256 \\
OM transformer & $d_{model}{=}128$, 2 layers, 4 heads, $d_{ff}{=}256$,
  dropout 0.05, history length 50 \\
\bottomrule
\end{tabular}
\end{table}

\paragraph{Zero-shot evaluation.} Checkpoints are evaluated against
\textsc{greedy} (the training opponent), \textsc{simple}, and
\textsc{oracle-stalker}, with seat-swap probability 0.5 and 1000 episodes
per seed. We report differences paired by training seed
($\Delta = \om - \classic$; the same seed index for both agents), split
by $n$-step configuration (pooled only where stated, as in the
belief-prior comparison), mean $\pm$ std over seeds.

\section{Results}
\paragraph{Training stability}
\label{sec:stability}

\begin{figure*}[t]
    \centering
    \includegraphics[width=\textwidth]{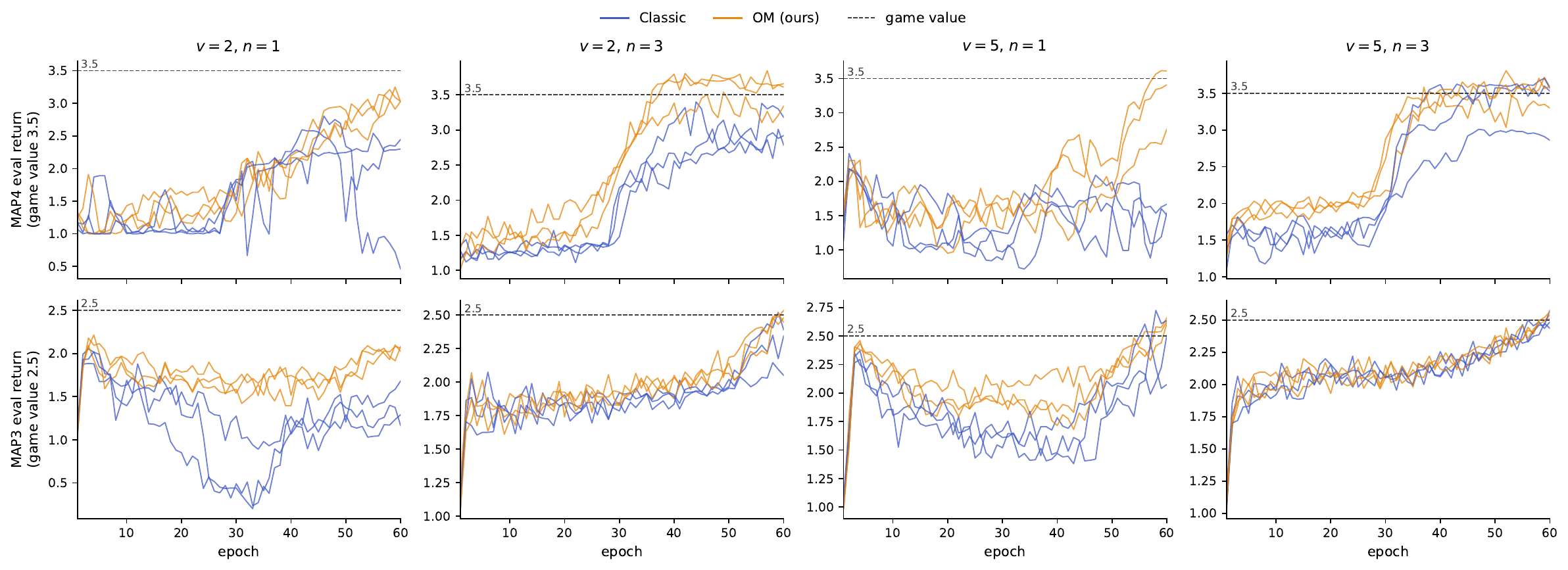}
    \caption{Per-epoch evaluation return against the training opponent, per
    seed: \classic{} (blue) vs.\ \om{} (orange), belief prior off. Top row
    \textsc{map4}, bottom row \textsc{map3}. \classic{} degenerates at
    $n{=}1$: on \textsc{map4} one $v{=}2$ seed collapses outright, all
    three $v{=}5$ seeds finish below $1.7$ of $3.5$, and on \textsc{map3}
    all seeds in the $v{=}2$ cell finish below $2.0$ of $2.5$ (thin fog
    recovers). \om{} never collapses on either map.}
    \label{fig:curves}
\end{figure*}

Plain DQN is unstable at $n{=}1$ (Figure~\ref{fig:curves}). On \textsc{map4}
one \classic{} seed collapses outright in $v{=}2$ and all three $v{=}5$ seeds
finish at 37--46\% of game value (1.3--1.6 of 3.5); on \textsc{map3} at thick fog all
$n{=}1$ seeds finish at 47--68\% of game value (1.17--1.71 of 2.5). \om{}
never collapses and never finishes below 80\% of game value on \textsc{map3}
(floor 2.03 of 2.5). Across all zero-shot batches on both maps, seed-level
standard deviation is 1.4--3.6$\times$ smaller for \om{} (\textsc{map4}:
0.16--0.31 vs.\ 0.50--0.79).

\paragraph{In-distribution performance}
\label{sec:indist}

\begin{figure*}[t]
    \centering
    \includegraphics[width=0.48\textwidth]{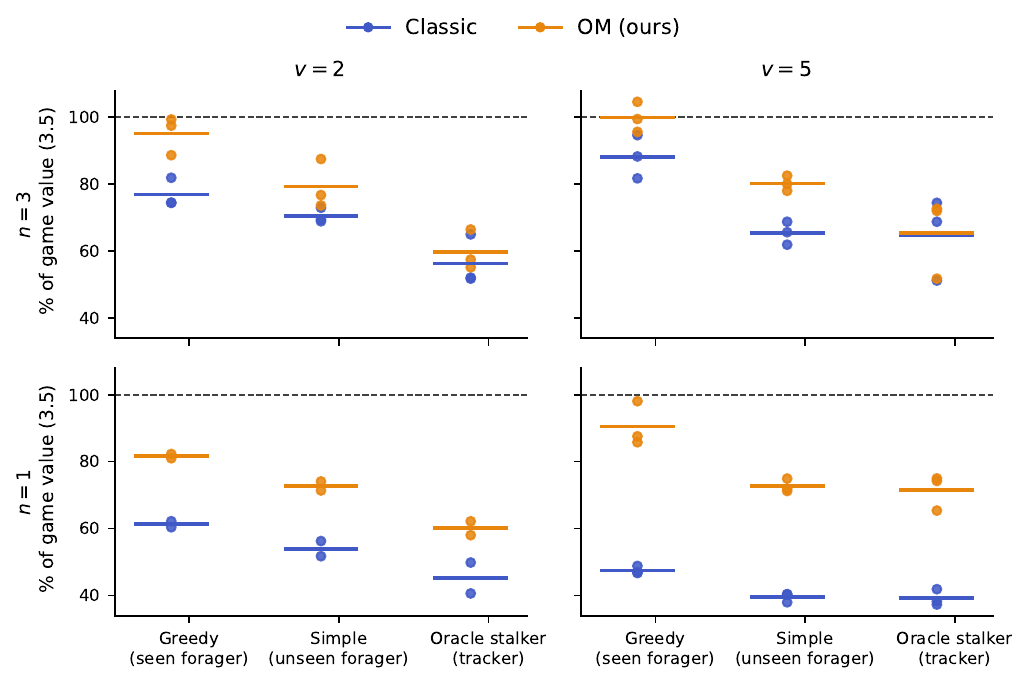}\hfill
    \includegraphics[width=0.48\textwidth]{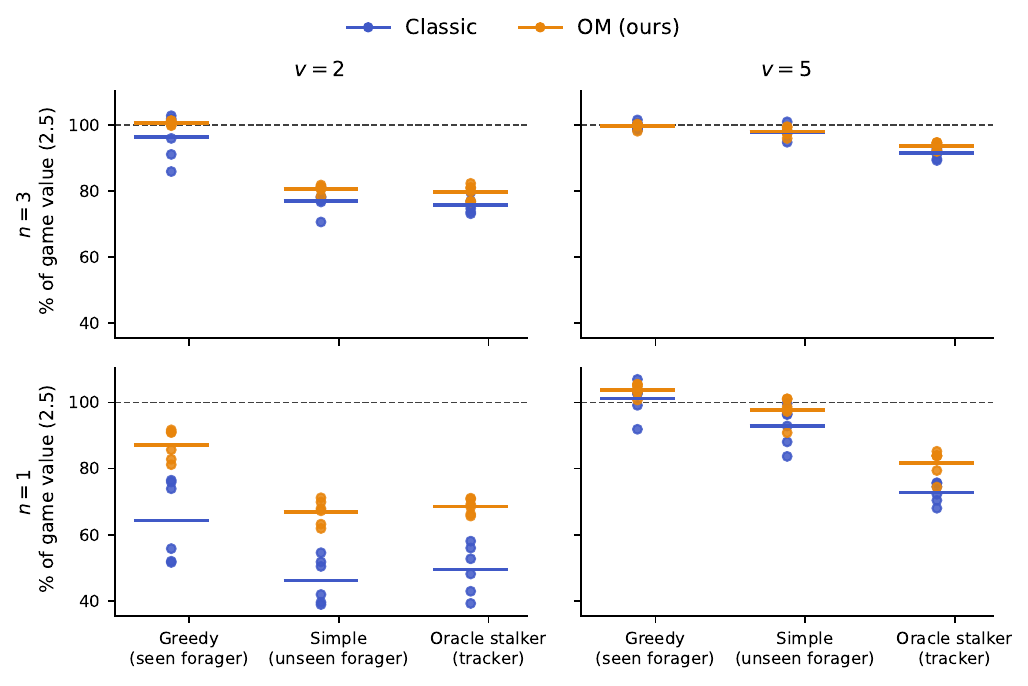}
    \caption{Zero-shot return as a share of the symmetric game value, by
    opponent and cell (dots: seeds; bar: mean; dashed: game value).
    Left: \textsc{map4} (prior off; collapsed \classic{} seed excluded).
    Right: \textsc{map3} (prior on/off pooled, 6 seeds per cell). Both agents
    exceed 100\% under thin fog (best-response exploitation of the suboptimal
    heuristic) and both saturate on the dense map wherever training succeeds.}
    \label{fig:value}
\end{figure*}

Against the training opponent (Figure~\ref{fig:value}), \om{} recovers 95\%
(\textsc{map4}, $v{=}2$) to 99.9\% ($v{=}5$) of the game value in well-trained
($n{=}3$) cells; \classic{} reaches 77--88\%. On \textsc{map3} both agents
reach 96--101\% at $n{=}3$, since the game is saturated. Under thin fog both
even exceed the game value on average (104\% for \om{}, 101\% for
\classic{}; individual seeds up to 107\%): a best response to the suboptimal
training heuristic. Thick fog separates them: in the hard cell ($v{=}2$,
$n{=}1$) \classic{} degenerates to 52--76\% per seed (64\% on average), because exploiting an opponent you
rarely see requires predicting it.

\paragraph{Opponent spectrum}
\label{sec:spectrum}

\begin{table}[t]
\centering
\caption{Paired zero-shot advantage $\Delta = \om - \classic$ (mean $\pm$ std
over seeds) by opponent, fog, and $n$-step horizon on \textsc{map4} (prior
off). $n{=}3$: well-trained cells;
$n{=}1$: cells where \classic{} training is unstable. Seen forager means that the learner has trained against the opponent; unseen forager means that the opponent is not in the learner's training set. Dense-map (\textsc{map3}) deltas are reported in the text and Table~\ref{tab:blindfold}.}
\label{tab:spectrum}
\small
\begin{tabular}{llcc}
\toprule
Opponent & Fog & $n{=}3$ & $n{=}1$ \\
\midrule
\textsc{greedy} (seen, forager) & $v{=}2$ & $+0.64 \pm 0.20$ & $+0.71 \pm 0.08^{\dagger}$ \\
\textsc{greedy} (seen, forager) & $v{=}5$ & $+0.41 \pm 0.07$ & $+1.50 \pm 0.26$ \\
\textsc{simple} (unseen, forager) & $v{=}2$ & $+0.31 \pm 0.31$ & $+0.66 \pm 0.05^{\dagger}$ \\
\textsc{simple} (unseen, forager) & $v{=}5$ & $+0.52 \pm 0.10$ & $+1.16 \pm 0.05$ \\
\textsc{oracle-stalker} (tracker) & $v{=}2$ & $+0.12 \pm 0.39$ & $+0.52 \pm 0.13^{\dagger}$ \\
\textsc{oracle-stalker} (tracker) & $v{=}5$ & $+0.02 \pm 0.09$ & $+1.13 \pm 0.27$ \\
\bottomrule
\end{tabular}
\\[2pt]
{\footnotesize $^{\dagger}$Excluding one collapsed \classic{} seed (see
Limitations); the collapse is reported as a finding.}
\end{table}

The advantage transfers across forager types, seen or unseen, and in
well-trained scenarios disappears against the tracker
(Table~\ref{tab:spectrum}). \textsc{simple} is unseen by the learner, but
its profile appears in the OM's offline pre-training corpus
(Section~\ref{sec:pretrain}), so this tests the learned policy's transfer,
not the OM's. This fits an information argument. A forager's
intent is a fact about the environment: its trajectory reveals it, and the
learner's fogged observations do not contain it. A tracker's intent is the
learner itself, so predicting it is redundant with self-knowledge. Tracking is
also costly for the tracker: total extraction drops from ${\sim}7$ goals
(against foragers) to ${\sim}3.5$, as the tracker spends roughly 2.2 goals of
its own foraging to remove about 1 from the learner.

On the dense map, pooling prior on and off, the paired advantage concentrates
in one hard cell ($v{=}2$, $n{=}1$): $+0.57$ vs.\
\textsc{greedy}, $+0.52$ vs.\ \textsc{simple}, $+0.48$ vs.\ the tracker.
Wherever \classic{} trains, the map saturates and the advantage falls to at
most $+0.13$ against foragers and $+0.23$ against the tracker. The tracker at
the hard cell is our mechanism probe: the advantage survives ($+0.48$), and
blinding the heatmap leaves it nearly intact ($+0.44$,
Table~\ref{tab:blindfold}). Against a tracker, the gap is training-time
stabilisation, not inference.

\paragraph{Ablating the inference channel}
\label{sec:blindfold}

\begin{table}[t]
\centering
\caption{Blindfold ablation: heatmap zeroed at deployment. Mean \om{}
return over seeds; $\Delta = \om{} - \classic{}$ paired by training seed,
zero-shot. \textsc{map4}:
prior off, $n$-split; \textsc{map3}: $n{=}1$, prior on/off pooled, 6 seeds;
vs.\ the opponent shown.}
\label{tab:blindfold}
\small
\setlength{\tabcolsep}{3.5pt}
\begin{tabular}{llcccc}
\toprule
Map / opponent & Cell & \om{} & \om{} blind & $\Delta$ & $\Delta$ blind \\
\midrule
\textsc{map4}, \textsc{greedy} & $v{=}2$, $n{=}3$ & 3.33 & 2.23 & $+0.64$ & $-0.47$ \\
\textsc{map4}, \textsc{greedy} & $v{=}2$, $n{=}1$ & 2.86 & 2.43 & $+0.71^{\dagger}$ & $+0.28^{\dagger}$ \\
\textsc{map4}, \textsc{simple} & $v{=}2$, $n{=}3$ & 2.77 & 2.10 & $+0.31$ & $-0.37$ \\
\textsc{map4}, \textsc{simple} & $v{=}2$, $n{=}1$ & 2.55 & 2.31 & $+0.66^{\dagger}$ & $+0.42^{\dagger}$ \\
\textsc{map4}, \textsc{oracle-stalker} & $v{=}2$, $n{=}3$ & 2.09 & 1.71 & $+0.12$ & $-0.26$ \\
\textsc{map4}, \textsc{oracle-stalker} & $v{=}2$, $n{=}1$ & 2.10 & 2.10 & $+0.52^{\dagger}$ & $+0.52^{\dagger}$ \\
\textsc{map4}, \textsc{greedy} & $v{=}5$, $n{=}3$ & 3.50 & 3.02 & $+0.41$ & $-0.07$ \\
\textsc{map4}, \textsc{greedy} & $v{=}5$, $n{=}1$ & 3.17 & 2.50 & $+1.50$ & $+0.84$ \\
\textsc{map4}, \textsc{simple} & $v{=}5$, $n{=}3$ & 2.81 & 2.57 & $+0.52$ & $+0.28$ \\
\textsc{map4}, \textsc{simple} & $v{=}5$, $n{=}1$ & 2.54 & 2.31 & $+1.16$ & $+0.92$ \\
\textsc{map4}, \textsc{oracle-stalker} & $v{=}5$, $n{=}3$ & 2.29 & 2.17 & $+0.02$ & $-0.10$ \\
\textsc{map4}, \textsc{oracle-stalker} & $v{=}5$, $n{=}1$ & 2.50 & 2.24 & $+1.13$ & $+0.87$ \\
\midrule
\textsc{map3}, \textsc{greedy} & $v{=}2$ & 2.18 & 2.13 & $+0.57$ & $+0.53$ \\
\textsc{map3}, \textsc{simple} & $v{=}2$ & 1.67 & 1.64 & $+0.52$ & $+0.48$ \\
\textsc{map3}, \textsc{oracle-stalker} & $v{=}2$ & 1.71 & 1.68 & $+0.48$ & $+0.44$ \\
\textsc{map3}, \textsc{greedy} & $v{=}5$ & 2.60 & 2.44 & $+0.07$ & $-0.09$ \\
\textsc{map3}, \textsc{simple} & $v{=}5$ & 2.44 & 2.29 & $+0.12$ & $-0.03$ \\
\textsc{map3}, \textsc{oracle-stalker} & $v{=}5$ & 2.04 & 2.03 & $+0.23$ & $+0.21$ \\
\bottomrule
\end{tabular}

{\footnotesize $^{\dagger}$Row excludes one collapsed \classic{} seed (see
Limitations); the collapse is reported as a finding.}
\end{table}

Table~\ref{tab:blindfold} completes the grid: all three opponents, both fog
levels, and both returns configurations, with the paired advantage computed
within the same runs. The picture splits along training difficulty. In the
well-trained cells ($n{=}3$), blinding erases the entire advantage in five of
six cells: against the training opponent at thick fog $+0.64 \to -0.47$,
against the unseen forager $+0.31 \to -0.37$. The one exception keeps just
over half ($+0.52 \to +0.28$, unseen forager, thin fog). Where training succeeds,
the advantage is inference. In the unstable cells ($n{=}1$), an edge survives
blinding against every opponent ($+0.28$ to $+0.92$ over \classic{},
excluding the collapsed seed at thick fog): stabilisation that does not
require the heatmap at deployment.

Reliance on the channel is strongest in distribution and in fog. The largest
drop is against the training opponent at thick fog ($3.33 \to 2.23$, leaving
\om{} half a goal below \classic{}), with a worst per-seed drop of $1.81$.
Across the well-trained cells the drop shrinks as fog
thins ($1.10 \to 0.48$ goals against the training opponent): the policy
consults the heatmap in proportion to how much its own observations leave
unknown. The unstable cells show no consistent fog scaling, so their
stabilisation is fog-independent even where their inference is not. Blinding
never collapses an \om{} seed (per-seed returns stay above $1.5$), so the
drop mixes information loss with input corruption (see
Limitations). \emph{Where training is hard, the
opponent model contributes stabilisation; where training is easy, it
contributes inference.}

The tracker rows tell the same story on the sparse map. In the unstable
cells the gap survives blinding almost fully ($+0.52 \to +0.52$ at thick
fog, $+0.87$ of $+1.13$ at thin fog): stabilisation. In the well-trained
cells the advantage is small, but its blindfold response still tracks fog:
near zero either way at thin fog ($+0.02 \to -0.10$), yet at thick fog the
channel is still consulted ($+0.12 \to -0.26$, a $0.37$-goal drop for \om{}
itself); the heatmap there carries not the target (the learner knows its own
position) but the tracker's approach under fog. An opponent whose intent is
about the learner rather than the environment
yields stabilisation, not inference.

On the dense map, blinding is nearly neutral at thick fog (at most $0.05$
goals in every $v{=}2$ cell), so the whole $+0.5$ advantage there is
stabilisation. The largest drops come at thin fog ($-0.16$ against the
training opponent at $v{=}5$) and flip the paired advantage slightly
negative ($-0.09$). With the game saturated there is little for inference
to add. The fog scaling of inference therefore needs a cell that is not
saturated, which the dense map lacks: itself a density effect.

\paragraph{Resource density}
\label{sec:density}

\begin{table}[t]
\centering
\caption{Dense \textsc{map3} (game value 2.5 by symmetry): epoch-60 eval
return vs.\ the training opponent as \% of game value, mean over 3 seeds per
cell. $\Delta$: paired \om$-$\classic{} in goals.}
\label{tab:density}
\small
\begin{tabular}{llcccc}
\toprule
Fog & $n$ & Prior & \classic{} & \om{} & $\Delta$ \\
\midrule
$v{=}2$ & 1 & off & 55.1 & 82.3 & $+0.68$ \\
$v{=}2$ & 1 & on  & 65.8 & 89.4 & $+0.59$ \\
$v{=}2$ & 3 & off & 90.5 & 99.5 & $+0.23$ \\
$v{=}2$ & 3 & on  & 98.4 & 99.8 & $+0.04$ \\
$v{=}5$ & 1 & off & 96.2 & 105.3 & $+0.23$ \\
$v{=}5$ & 1 & on  & 104.0 & 106.9 & $+0.07$ \\
$v{=}5$ & 3 & off & 99.9 & 101.5 & $+0.04$ \\
$v{=}5$ & 3 & on  & 97.8 & 99.6 & $+0.05$ \\
\bottomrule
\end{tabular}
\end{table}

Table~\ref{tab:density} completes the picture. Wherever \classic{} trains
successfully ($\geq 90\%$ of value), \om{} also sits at the ceiling and the
advantage is at most $0.23$ goals. On a dense map a lost contest costs little
because another goal is nearby, so predicting the opponent's target is worth
little: the advantage vanishes by saturation, not by failure. The stability
effect replicates and depends on fog. At thick fog all six \classic{} $n{=}1$
seeds fall below $2.0$ of $2.5$ (minimum $1.17$) while \om{} never drops below
$2.03$; at thin fog even \classic{} $n{=}1$ recovers (minimum $2.08$). Across
both maps, the advantage shrinks with the difficulty reducers that stabilise
\classic{} training (denser map, longer $n$-step horizon), from $+0.5$--$1.5$
goals in the sparse map's unstable cells to $+0.04$ on the dense map under
thin fog. Fog is the exception on the sparse map: at $n{=}1$ the gap widens
as fog thins ($+0.71 \to +1.50$ vs.\ the training opponent), because \om{}
improves with visibility ($2.84 \to 3.17$) while \classic{} does not
($2.1 \to 1.7$ goals excluding the collapsed seed).

\paragraph{Belief-map prior}
\label{sec:prior}

On \textsc{map3} the prior is neutral for \om{}: the largest cell-level return
difference between $p{=}1$ and $p{=}0$ is $0.18$ goals, and its effect on
inference quality is small and inconsistent in sign ($-0.28$ to
$+0.26$). It partially rescues \classic{} where TD(0) fails: at the hard cell
($v{=}2$, $n{=}1$; zero-shot vs.\ \textsc{greedy}), \classic{} rises from
53\% to 75\% of game value with the prior, while \om{} remains highest
(91\%). Table~\ref{tab:density} shows the same
direction with smaller gaps (55.1\% $\to$ 65.8\%, \om{} 82.3--89.4\%).

The \textsc{map4} replication turns this into a
boundary condition. At thin fog the prior lifts the baseline (\classic{} vs.\
the training opponent: 68\% $\to$ 84\% of game value) while \om{} gives up
$0.3$--$0.6$ goals. The paired advantage, pooled over the $n$-split for this comparison, thus
shrinks from $+0.90$/$+0.96$ to
$+0.60$/$+0.04$ goals (thick/thin fog) and nearly vanishes
against the tracker (from $+0.51$/$+0.58$ to $+0.01$/$+0.24$). Part of the prior-free
edge was an information-starved baseline: a positional belief substitutes for
subgoal inference at thin fog, but not at thick fog, where inference keeps a
$+0.6$ goal edge over the baseline with the prior. The prior is a safe default
for the baseline; the opponent model buys the thick-fog remainder.

\section{Limitations}
Three seeds per configuration. We report per-seed
  strip plots and never pool across $n$-step configurations (except where
  stated, as in the belief-prior comparison), and several
  cells have seed-level standard deviations comparable to their means and
  should be read as non-significant. One
  \classic{} seed on \textsc{map4} collapsed during training and is
  excluded from paired statistics there, with the training curve cited as
  justification. Collapse frequency is itself reported. All results are best responses to
  a heuristic. In preliminary experiments, NFSP-style
  self-play~\cite{DBLP:journals/corr/HeinrichS16} did not
  converge in the FOSG domain, so we make no MARL equilibrium claims.
  Published opponent-modeling methods
  assume different observation spaces (full state, action histories, or
  population priors). None infers spatial opponent subgoals from partial
  observations in this setting, so we compare against \classic{}, whose
  Q-network matches the \om{} Q-network on observation and belief channels
  and on capacity, differing only in the heatmap input, although that
  heatmap summarises an observation history \classic{} does not see (see
  Conclusion). We fix the learner to value-based DQN
  methods because the claim concerns TD(0) stability
  under partial observability. Policy-gradient learners would change the
  failure mode rather than test it. Zeroing the heatmap is an input distribution shift. The drop measures information loss plus input corruption, not information loss alone. And lastly, this work focuses on grid foraging only, scale is future
  work.

\section{Conclusion and Future Work}
Opponent-goal inference under partial observability is not a uniform
advantage: its value is set by visibility (fog thickness), by whether the opponent's
intent is about the environment or about the learner, and by resource
density. Its most robust benefit is making DQN training stable
where plain TD(0) fails. We attribute the stabilisation to the module as a whole. Whether it stems from subgoal inference or from history-derived structured input is not isolated here. The blindfold ablation offers partial evidence: zeroing the heatmap, the only input beyond the channels shared with \classic{}, costs the most deployment performance in the cells where the information argument says inference matters most (in distribution and at thick fog on the sparse map), consistent with the inference carrying information rather than acting as generic structured input. Future work scales these mechanisms to
roadmap graphs of real-world maps, where foraging under fog
becomes a multi-agent travelling-salesman problem under partial
observability, in both competitive and cooperative settings, and
replaces self-play with policy-space or population methods for
equilibrium approximation.

\section{Acknowledgements}
I would like to thank my supervisor, Ing. Rostislav Horčík, Ph.D., for his invaluable guidance and support throughout this research. The authors acknowledge the support of the OP VVV ``Research Center for Informatics''. The access to the computational infrastructure of the OP VVV ``Research Center for Informatics'' is also gratefully acknowledged.

\clearpage
\bibliographystyle{ACM-Reference-Format}
\bibliography{sample}

\balance
\end{document}